\documentstyle[psfig,epsf,12pt]{article}
\catcode`\@=11
\long\def\@makefntext#1{ 
\protect\noindent \hbox to 3.2pt {\hskip-.9pt
$^{{\ninerm\@thefnmark}}$\hfil}#1\hfill} 

\def\thefootnote{\fnsymbol{footnote}}
 \def\@makefnmark{\hbox to 0pt{$^{\@thefnmark}$\hss}}  

\def\ps@myheadings{\let\@mkboth\@gobbletwo
\def\@oddhead{\hbox{} 
\rightmark\hfil\ninerm\thepage}
\def\@oddfoot{}\def\@evenhead{\ninerm\thepage\hfil 
\leftmark\hbox{}}\def\@evenfoot{}
\def\sectionmark##1{}\def\subsectionmark##1{}}

\begin{document}

\newcommand{\be}{\begin{equation}}
\newcommand{\ee}{\end{equation}}
\newcommand{\bc}{\begin{center}}
\newcommand{\ec}{\end{center}}
\newcommand{\ba}{\begin{array}}
\newcommand{\ea}{\end{array}}
\newcommand{\plb}[1]{Phys. Lett. {\bf #1}}
\newcommand{\npb}[1]{Nucl. Phys. {\bf #1}}
\newcommand{\prb}[1]{Phys. Rev. {\bf #1}}
\newcommand{\prl}[1]{Phys. Rev. Lett. {\bf #1}}
\newcommand{\cD}{{\cal D}}


\newcommand{\symbolfootnote}{\renewcommand{\thefootnote}
	{\fnsymbol{footnote}}}
\renewcommand{\thefootnote}{\fnsymbol{footnote}}
\newcommand{\alphfootnote}
	{\setcounter{footnote}{0}
	 \renewcommand{\thefootnote}{\sevenrm\alph{footnote}}}

\newcounter{sectionc}\newcounter{subsectionc}\newcounter{subsubsectionc}
\renewcommand{\section}[1] {\vspace{0.6cm}\addtocounter{sectionc}{1}
\setcounter{subsectionc}{0}\setcounter{subsubsectionc}{0}\noindent
	{\bf\thesectionc. #1}\par\vspace{0.4cm}}
\renewcommand{\subsection}[1] {\vspace{0.6cm}\addtocounter{subsectionc}{1}
	\setcounter{subsubsectionc}{0}\noindent
	{\it\thesectionc.\thesubsectionc. #1}\par\vspace{0.4cm}}
\renewcommand{\subsubsection}[1]
{\vspace{0.6cm}\addtocounter{subsubsectionc}{1}
	\noindent {\rm\thesectionc.\thesubsectionc.\thesubsubsectionc.
	#1}\par\vspace{0.4cm}}
\newcommand{\nonumsection}[1] {\vspace{0.6cm}\noindent{\bf #1}
	\par\vspace{0.4cm}}

\newcounter{appendixc}
\newcounter{subappendixc}[appendixc]
\newcounter{subsubappendixc}[subappendixc]
\renewcommand{\thesubappendixc}{\Alph{appendixc}.\arabic{subappendixc}}
\renewcommand{\thesubsubappendixc}
	{\Alph{appendixc}.\arabic{subappendixc}.\arabic{subsubappendixc}}

\renewcommand{\appendix}[1] {\vspace{0.6cm}
        \refstepcounter{appendixc}
        \setcounter{figure}{0}
        \setcounter{table}{0}
        \setcounter{equation}{0}
        \renewcommand{\thefigure}{\Alph{appendixc}.\arabic{figure}}
        \renewcommand{\thetable}{\Alph{appendixc}.\arabic{table}}
        \renewcommand{\theappendixc}{\Alph{appendixc}}
        \renewcommand{\theequation}{\Alph{appendixc}.\arabic{equation}}
        \noindent{\bf Appendix \theappendixc #1}\par\vspace{0.4cm}}
\newcommand{\subappendix}[1] {\vspace{0.6cm}
        \refstepcounter{subappendixc}
        \noindent{\bf Appendix \thesubappendixc. #1}\par\vspace{0.4cm}}
\newcommand{\subsubappendix}[1] {\vspace{0.6cm}
        \refstepcounter{subsubappendixc}
        \noindent{\it Appendix \thesubsubappendixc. #1}
	\par\vspace{0.4cm}}

\def\abstracts#1{{
	\centering{\begin{minipage}{30pc}\tenrm\baselineskip=12pt\noindent
	\centerline{\tenrm ABSTRACT}\vspace{0.3cm}
	\parindent=0pt #1
	\end{minipage} }\par}}

\newcommand{\bibit}{\it}
\newcommand{\bibbf}{\bf}
\renewenvironment{thebibliography}[1]
	{\begin{list}{\arabic{enumi}.}
	{\usecounter{enumi}\setlength{\parsep}{0pt}
\setlength{\leftmargin 1.25cm}{\rightmargin 0pt}
	 \setlength{\itemsep}{0pt} \settowidth
	{\labelwidth}{#1.}\sloppy}}{\end{list}}

\topsep=0in\parsep=0in\itemsep=0in
\parindent=1.5pc

\newcounter{itemlistc}
\newcounter{romanlistc}
\newcounter{alphlistc}
\newcounter{arabiclistc}
\newenvironment{itemlist}
    	{\setcounter{itemlistc}{0}
	 \begin{list}{$\bullet$}
	{\usecounter{itemlistc}
	 \setlength{\parsep}{0pt}
	 \setlength{\itemsep}{0pt}}}{\end{list}}

\newenvironment{romanlist}
	{\setcounter{romanlistc}{0}
	 \begin{list}{$($\roman{romanlistc}$)$}
	{\usecounter{romanlistc}
	 \setlength{\parsep}{0pt}
	 \setlength{\itemsep}{0pt}}}{\end{list}}

\newenvironment{alphlist}
	{\setcounter{alphlistc}{0}
	 \begin{list}{$($\alph{alphlistc}$)$}
	{\usecounter{alphlistc}
	 \setlength{\parsep}{0pt}
	 \setlength{\itemsep}{0pt}}}{\end{list}}

\newenvironment{arabiclist}
	{\setcounter{arabiclistc}{0}
	 \begin{list}{\arabic{arabiclistc}}
	{\usecounter{arabiclistc}
	 \setlength{\parsep}{0pt}
	 \setlength{\itemsep}{0pt}}}{\end{list}}

\newcommand{\fcaption}[1]{
        \refstepcounter{figure}
        \setbox\@tempboxa = \hbox{\tenrm Fig.~\thefigure. #1}
        \ifdim \wd\@tempboxa > 6in
           {\begin{center}
        \parbox{6in}{\tenrm\baselineskip=12pt Fig.~\thefigure. #1 }
            \end{center}}
        \else
             {\begin{center}
             {\tenrm Fig.~\thefigure. #1}
              \end{center}}
        \fi}

\newcommand{\tcaption}[1]{
        \refstepcounter{table}
        \setbox\@tempboxa = \hbox{\tenrm Table~\thetable. #1}
        \ifdim \wd\@tempboxa > 6in
           {\begin{center}
        \parbox{6in}{\tenrm\baselineskip=12pt Table~\thetable. #1 }
            \end{center}}
        \else
             {\begin{center}
             {\tenrm Table~\thetable. #1}
              \end{center}}
        \fi}

\def\@citex[#1]#2{\if@filesw\immediate\write\@auxout
	{\string\citation{#2}}\fi
\def\@citea{}\@cite{\@for\@citeb:=#2\do
	{\@citea\def\@citea{,}\@ifundefined
	{b@\@citeb}{{\bf ?}\@warning
	{Citation `\@citeb' on page \thepage \space undefined}}
	{\csname b@\@citeb\endcsname}}}{#1}}

\newif\if@cghi
\def\cite{\@cghitrue\@ifnextchar [{\@tempswatrue
	\@citex}{\@tempswafalse\@citex[]}}
\def\citelow{\@cghifalse\@ifnextchar [{\@tempswatrue
	\@citex}{\@tempswafalse\@citex[]}}
\def\@cite#1#2{{$\null^{#1}$\if@tempswa\typeout
	{IJCGA warning: optional citation argument
	ignored: `#2'} \fi}}
\newcommand{\citeup}{\cite}

\def\fnm#1{$^{\mbox{\scriptsize #1}}$}
\def\fnt#1#2{\footnotetext{\kern-.3em
	{$^{\mbox{\sevenrm #1}}$}{#2}}}

\font\twelvebf=cmbx10 scaled\magstep 1
\font\twelverm=cmr10 scaled\magstep 1
\font\twelveit=cmti10 scaled\magstep 1
\font\elevenbfit=cmbxti10 scaled\magstephalf
\font\elevenbf=cmbx10 scaled\magstephalf
\font\elevenrm=cmr10 scaled\magstephalf
\font\elevenit=cmti10 scaled\magstephalf
\font\bfit=cmbxti10
\font\tenbf=cmbx10
\font\tenrm=cmr10
\font\tenit=cmti10
\font\ninebf=cmbx9
\font\ninerm=cmr9
\font\nineit=cmti9
\font\eightbf=cmbx8
\font\eightrm=cmr8
\font\eightit=cmti8


\centerline{\twelvebf Temperature dependence of the response function
of hot nuclear matter }
\vspace{0.8cm}
\centerline{\tenrm Abdellatif Abada$^1$ and Dominique Vautherin$^2$}
\baselineskip=13pt
\centerline{\tenit $^1$~Theoretical Physics Group, Department of Physics and
Astronomy,}
\centerline{\tenit University of Manchester, Manchester M13 9PL, UK}
\centerline{\tenit $^2$~Division de Physique Th\'eorique \footnote{Unit\'e de
Recherche des Universit\'es Paris XI et Paris VI associ\'ee au CNRS},
Institut de Physique Nucl\'eaire }
\baselineskip=12pt
\centerline{\tenit Orsay Cedex, F-91406, France}
\vspace{0.9cm}
\abstracts{
The description of collective motion in nuclei at finite temperature
using the framework of the random phase approximation is discussed. We
focus on the special case of the isovector response function of hot nuclear
matter using various effective Skyrme interactions.}
\vfil

\vspace{0.3cm}
\hfill{MC/TH 95/20}
\vskip 0.3cm

\twelverm   
\baselineskip=14pt

\section{Introduction}

The purpose of the present notes is to investigate the response
function of hot nuclear matter. We use the framework of time dependent
mean field theory with the standard Skyrme type
effective interactions which have
provided in the past good descriptions of average nuclear properties.
For such interactions analytic expressions for the response function
are derived. From these analytic formulae we show that one can
gain an understanding of the temperature dependance of the giant
dipole resonance in finite nuclei. Although effects which are beyond
the framework of linear response theory are important to account for
the full width of the giant dipole state (e.g. two particle two hole
configurations) many features of these resonances are known to be
readily understood at this level already
\cite{RING,SOMMERMAN,SAGAWA,VAUVIN,NICOLE}. In the following we
show that zero sound type modes become less collective as temperature
rises and sometimes disappear at temperatures of a few
MeV. We also show that for effective interactions with an
effective mass close to unity collective effects are small already at
zero temperature and depend weakly on temperature. We will thus argue
that the temperature dependance of the giant dipole resonance provides
valuable information on the effective interaction.

\section{ Linear response theory in hot nuclear matter }
\setcounter{equation}{0}

When an external field of the form
\be \label{1e1}
V_{\rm ext}= \epsilon \tau_3 e^{i({\bf q}. {\bf r}- (\omega + i\eta) t},
\ee
is applied to nuclear matter, it induces a density change between neutron and
proton densities which is of the form
\be \label{1e2}
\delta \rho ({\bf r}, t)= \rho_n ({\bf r}, t)- \rho({\bf r},t)=
\eta e^{i({\bf q}. {\bf r}- (\omega + i\eta) t}.
\ee
The ratio of the amplitude $\alpha$ of the density oscillations to the external
field strength is the response function (also called dynamic polarizability)
\be \label{1e3}
\Pi (\omega, q)= \frac{\alpha}{\epsilon}
\ee
In the case of Skyrme type forces it is possible to find analytic expressions
for the response function of nuclear matter at finite temperature
\cite{ABADA,NAVARRO}.

Calculations are particularly simple in the case of a simplified
Skyrme interaction
\be \label{17e3}
v=t_0 (1+x_0 P_\sigma) \delta ({\bf r}_1 - {\bf r}_2)
+ t_3 \delta ({\bf r}_1 - {\bf r}_2) \delta ({\bf r}_2 - {\bf r}_3).
\ee
In this particular case one has the simple formula
\be \label{17e10}
\Pi(\omega, {\bf q})= \frac{\Pi_0 (\omega, {\bf q}) }
{1- V_0 \Pi_0 (\omega, {\bf q}) },
\ee
where $\Pi_0$ is the unperturbed polarization propagator \cite{WAL71},
also called Lindhard function
\be \label{17e11}
\Pi_0 (\omega, {\bf q}) =
\frac{2}{(2 \pi)^3} \int d {\bf k}
\frac{ f({\bf k}+ {\bf q})- f({\bf k})}
{\omega+i \eta- \epsilon({\bf k}) + \epsilon({\bf k}+ {\bf q} ) }.
\ee
and where
\be \label{17e5}
V_0= -t_0 (x_0+ 1/2 )/2 -t_3 \rho_0/8.
\ee
For a given momentum ${\bf q}$ we thus see that there is a resonant response
when the frequency $\omega$ of the external field corresponds to a zero
in the denominator i.e. when
\be \label{17e12}
1=V_0 \Pi_0(\omega, {\bf q}).
\ee
The real part of $\omega$ determines the energy of the collective mode and
its imaginary part the lifetime \cite{WAL71}.

The previous formulae correspond to the special case of a simplified
Skyrme force. They can however
be generalized to the case of a full Skyrme force. Such an interaction contains
a density independent part $v_{12}$ and a density dependent part $v_{123}$. The
density independent part is parametrized in momentum space as
\be \label{1e5}
<{\bf k}| v_{12}|{\bf k}'>= t_0(1+x_0 P_{\sigma})+ \frac{t_1}{2} (1+x_1
P_{\sigma}) (k^2 +k'^2) + t_2 (1+x_2 P_{\sigma}) {\bf k}.{\bf k}'.
\ee
The density dependent part in ordinary space is
\be \label{1e6}
v_{123}= \frac{t_3}{6} (1+ x_3 P_{\sigma}) \delta ({\bf r}_1- {\bf r}_2)
\rho^{\sigma} (\frac{{\bf r}_1+{\bf r}_2}{2}).
\ee
For such an interaction the response function is given by
\begin{equation} \label{15}
\Pi(\omega, {\bf q}) = \frac{\Pi_{0}
(\omega,{\bf q}) }{1- \bar{V_0} \Pi_{0}(\omega, {\bf q})
- 2 V_1 \Pi_2 (\omega, {\bf q})
- V_1^2 \Pi_2^2 (\omega, {\bf q}) - V_1^2 \Pi_4 \Pi_0  }.
\end{equation}
In this equation $\Pi_2$ and $\Pi_4$ are generalized Lindhard functions defined
by
\begin{equation} \label{16}
\Pi_{2N}(\omega, {\bf q})  = \frac{4}{ (2 \pi)^3}
\int \hbox{d}^3 k  \frac
{ f({\bf k} + {\bf q})- f( {\bf k})  }
{\omega +i \eta-  \epsilon ({\bf k}) + \epsilon({\bf k}+ {\bf q})  }
 \left( {\bf k} .({\bf k +q }) \right) ^N ~,
\end{equation}
while $V_0$ and $V_2$ are linear combinations of the parameters of the Skyrme
interaction. Explicitely one has
\begin{equation} \label{7} \begin{array}{ll}
V_0=&{\displaystyle
 - \frac{t_0}{2} \left(x_0+ \frac{1}{2} \right) - \frac{t_3}{12}\left(
x_3 + \frac{1}{2} \right) \rho_0^{\alpha} } \\
&{\displaystyle - \frac{q^2}{16} ( 3 t_1( 1+2 x_1) + t_2 ( 1+ 2 x_2) ) } \\
&{\displaystyle - \left( \frac{ m^* \omega}{q} \right)^2 \frac{ 2V_1}
{ 1- 2 V_1 m^*\rho_0 },
}
\end{array}
\end{equation}
where $m^*$ is the effective mass and $\rho_0$ the saturation density of
nuclear matter, while $V_1$ is:
\be \label{1e8}
{\displaystyle V_1= \frac{1}{16} ( t_2( 1+ 2 x_2) -  t_1( 1 + 2 x_1) )}
\ee
In reference \cite{BRAGHIN} a discussion of the properties of the response
function in the case of a simplified Skyrme interaction was provided. It was
found that for a momentum transfer $q$=0.23 fm$^{-1}$ (a value which
corresponds to the giant dipole resonance in lead- 208 in the Steinwedel-
Jenssen model) the response  function exhibits a sharp zero sound type peak at
zero temperature. In contrast when temperature rises the smearing of the
Lindhard function was found to make the peak much weaker and in fact nearly
disappear for temperatures of a few MeV. Whether this conclusion holds for
the more realistic Skyrme forces is a question which was investigated in
reference \cite{ABADA}. Some results of this study are shown in Figure 1.

\begin{figure}[t]
\centerline{\psfig{figure=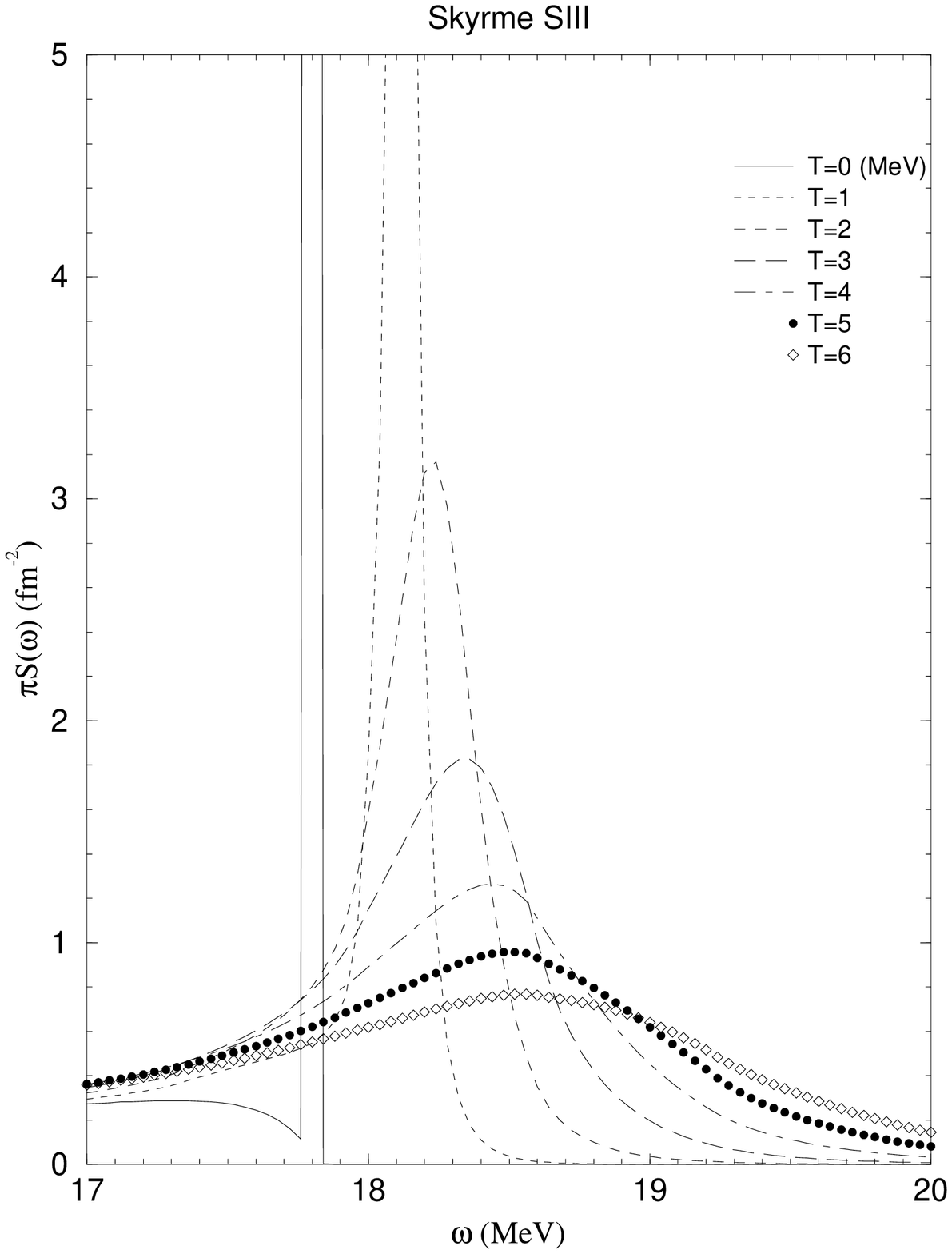,height=7cm,width=6cm}
\hskip 1.5cm      \psfig{figure=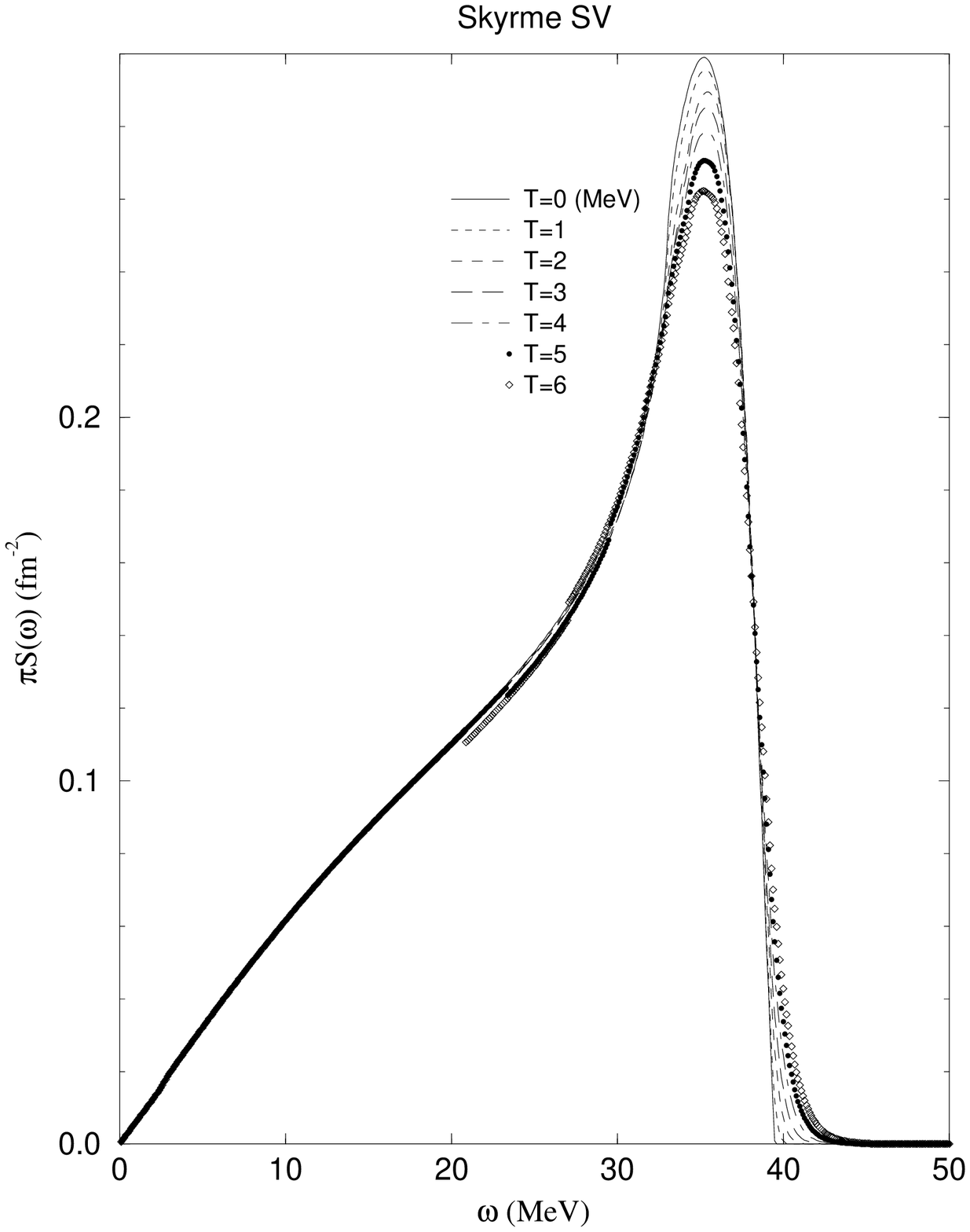,height=7cm,width=6cm}}
\vskip 1cm
\fcaption{Distribution of strength as a function of the frequency $\omega$
for a momentum transfer $q$=0.23 fm$^{-1}$ in the case of interactions SIII
and SV}
\end{figure}

Fig.1a (left side) gives the strength
\begin{equation} \label{30}
S(\omega)= -\frac{1}{\pi} \Im m\Pi(\omega,q).
\end{equation}
as a function of the frequency $\omega$ for a momentum transfer $q$= 0.23
fm$^{-1}$ in the case of the Skyrme force SIII while Fig.1b (right side)
corresponds to
interaction SV. It can be seen that for interaction SIII collective effects
disappear as soon as temperature reaches a few MeV while for SV only a weak
temperature dependance is observed. In reference \cite{ABADA} the difference
between these two cases was attributed to a difference between the effective
mass values for the two interactions. This difference shows that the
temperature dependance of the response function provides information on the
effective particle hole interaction. In this context it is worthwhile noting
that recent experimental studies of the giant dipole resonance found a
saturation of photon multiplicities at high excitation energy \cite{LEFAOU}.
This may be a signature of the loss of collectivity found at high temperature
in the present calculation.

\vskip 1cm
{\bf Acknowledgements}
One of us (D.V.)  wishes to thank Prof. Apolodor Raduta for his
invitation to present this lecture at the 1995 Predeal School on Nuclear
Structure.


\begin{thebibliography}{20}

\bibitem{RING} M. E. Faber, J. L. Egido and P. Ring, Phys. Leett. {\bf 127B}
(1983) 5
\bibitem{SOMMERMAN} M. Sommerman, Ann. of Phys. {\bf 151} (1983) 163
\bibitem{SAGAWA} H. Sagawa and G. F. Bertsch, Phys. Lett. {\bf 333B} (1994)
289
\bibitem{VAUVIN} D. Vautherin and N. Vinh Mau, Nucl. Phys. {\bf A422} (1984)
140
\bibitem{NICOLE} N. Vinh Mau, Nucl. Phys. {\bf A548} (1992) 381
\bibitem{ABADA} F. Braghin, D. Vautherin and A. Abada, Phys. Rev. C (1995),
in press
\bibitem{NAVARRO} E.S.Hernandez, J. Navarro, A. Polls and J. Ventura, Finite
Temperature RPA in Symmetric Nuclear Matter with Skyrme Interactions,
University
of Valencia preprint, July 27th, 1995
\bibitem{WAL71} A.L. Fetter and J.D. Walecka,
Quantum Theory of Many Particle Systems (Mc Graw-Hill, New York, 1971)
\bibitem{GIAI92} C. Garcia- Recio, J. Navarro, Nguyen van Giai and L.L Salcedo
Ann. Phys. {\bf 214} (1992) 293
\bibitem{BRAGHIN} F. Braghin and D. Vautherin, Phys. Lett. {\bf 333B} (1994)
289
\bibitem{LEFAOU} J. H. Le Faou {\it et al.}, Phys. Rev. Lett.
{\bf 72} (1994) 3321
\end{thebibliography}
\end{document}